\documentclass[a4paper,11pt]{article}

\usepackage{jheppub}
\numberwithin{equation}{section} \allowdisplaybreaks[4]

\newcommand{\bdm}{\begin{displaymath}}
\newcommand{\edm}{\end{displaymath}}
\newcommand{\beq}{\begin{equation}}
\newcommand{\eeq}{\end{equation}}
\newcommand{\bea}{\begin{eqnarray}}
\newcommand{\eea}{\end{eqnarray}}

\newcommand{\lt}{\left}
\newcommand{\rt}{\right}
\newcommand{\no}{\nonumber}

\newcommand{\ov}{\overline}
\newcommand{\eq}[1]{Eq.~(\ref{#1})}
\newcommand{\eqsand}[2]{Eqs.~(\ref{#1}) and (\ref{#2})}
\newcommand{\eqsto}[2]{Eqs.~(\ref{#1}) to (\ref{#2})}

\newcommand{\imag}{\mbox{Im\,}}
\newcommand{\real}{\mbox{Re\,}}

\newcommand{\Bbar}{\bar{B}}

\newcommand{\bbq}{\ensuremath{B_q\!-\!\Bbar{}_q\,}}

\newcommand{\bbmq}{\bbq\ mixing}

\newcommand{\bra}[1]{\ensuremath{\langle #1 |}}
\newcommand{\ket}[1]{\ensuremath{| #1 \rangle }}
\newcommand{\fig}[1]{Fig.~\ref{#1}}
\newcommand{\tab}[1]{Tab.~\ref{#1}}
\newcommand{\lqcd}{\Lambda_{\textit{\scriptsize{QCD}}}}

\newcommand{\dm}{\ensuremath{\Delta M}}
\newcommand{\dg}{\ensuremath{\Delta \Gamma}}
\newcommand{\epm}[2]{
 \raisebox{-0.5ex}{\shortstack[l]{$\scriptstyle+#1$\\$\scriptstyle-#2$}}}
\newcommand{\ds}{\displaystyle}

\def \be{\begin{equation}}
\def \ee{\end{equation}}

\catcode`@=11
\def\marginnote#1{}
\newcount\hour
\newcount\minute
\newtoks\amorpm
\hour=\time\divide\hour by60 \minute=\time{\multiply\hour by60
\global\advance\minute by-\hour}
\edef\standardtime{{\ifnum\hour<12 \global\amorpm={am}%
        \else\global\amorpm={pm}\advance\hour by-12 \fi
        \ifnum\hour=0 \hour=12 \fi
        \number\hour:\ifnum\minute<10 0\fi\number\minute\the\amorpm}}
\edef\militarytime{\number\hour:\ifnum\minute<10 0\fi\number\minute}
\def\draftlabel#1{{\@bsphack\if@filesw {\let\thepage\relax
   \xdef\@gtempa{\write\@auxout{\string
      \newlabel{#1}{{\@currentlabel}{\thepage}}}}}\@gtempa
   \if@nobreak \ifvmode\nobreak\fi\fi\fi\@esphack}
        \gdef\@eqnlabel{#1}}
\def\@eqnlabel{}
\def\@vacuum{}
\def\draftmarginnote#1{\marginpar{\raggedright\scriptsize\tt#1}}
\def\draft{\oddsidemargin 0.0truein
        \def\@oddfoot{\sl preliminary draft \hfil
        \rm\thepage\hfil\sl\today\quad\militarytime}
        \let\@evenfoot\@oddfoot \overfullrule 3pt
        \let\label=\draftlabel
        \let\marginnote=\draftmarginnote
   \def\@eqnnum{(\theequation)\rlap{\kern\marginparsep\tt\@eqnlabel}%
\global\let\@eqnlabel\@vacuum}  } \catcode`@=12

\title{\vskip-4.5cm{
    \begin{flushright}
      \small TTP25-045,  P3H-25-095 
    \end{flushright}} \vskip1.5cm \boldmath Decay matrix of $B$-$\bar B$ mixing:
      Mixing of dimension-seven operators into dimension-six operators
      under renormalization}

    \author{Artyom Hovhannisyan$^a$ and}
    \author{Ulrich Nierste$^b$}
    \emailAdd{artyom@yerphi.am}
      \emailAdd{ulrich.nierste@kit.edu}

\affiliation{$^{\,a}$A. I. Alikhanyan National Science
      Laboratory (Yerevan Physics Institute), Alikhanian Br. Str. 2,
      0036 Yerevan, Armenia}

\affiliation{$^{\,b}$Institut f{\"u}r Theoretische Teilchenphysik, Karlsruher
    Institut f{\"u}r Technologie, Wolfgang-Gaede-Str. 1, 76131 Karlsruhe, Germany} 

  \abstract{{The precise measurement of the width difference $\dg_s$
      among the mass eigenstates of the $B_s$-$\bar B_s$ system requires
      the calculation of the corresponding decay matrix to order
      $\alpha_s/m_b$. QCD corrections to power-suppressed terms in the
      Heavy Quark Expansion involve the renormalization of dimension-7
      four-quark operators for which no general methodology is available
      yet. In the $\ov{\rm MS}$ scheme one-loop 
      corrections to matrix elements of dimension-7 operators violate
      the power counting, but we find the responsible terms to be
      infrared-finite and show that they can be absorbed into finite
      counterterms proportional to dimension-6 operators. We calculate
      all these counterterms and subsequently verify the consistency of
      our results with hadronic matrix elements calculated in the limit
      of a large number $N_c$ of colours. The condition of correct power
      counting implies constraints on the possible definitions of
      evanescent operators.}}

\begin{document}
\maketitle
\section{Introduction}
A neutral $B_q$ meson mixes with its antimeson $\bar B_q$ leading to
a time evolution formula featuring  damped oscillations between these two
states: 
\begin{equation}
i \frac{d}{dt}
\left(
\begin{array}{c}
\ket{B_q(t)} \\ \ket{\bar{B}_q (t)}
\end{array}
\right)
=
\left( M^q - \frac{i}{2} \Gamma^q \right)
\left(
\begin{array}{c}
\ket{B_q(t)} \\ \ket{\bar{B}_q (t)} 
\end{array}
\right),\qquad \qquad q=d \mbox{ or } s,\label{sch}
\end{equation} 
with the hermitian $2\times 2$ mass and decay matrices $M^q$ and
$ \Gamma^q $, respectively. Relevant for \bbmq\ are the off-diagonal
elements $M_{12}^q=M_{21}^{q*}$ and $\Gamma_{12}^q=\Gamma_{21}^{q*}$.
The diagonalization of $M^q-i\Gamma^q/2$ yields the mass eigenstates
$B_L^q$ and $B_H^q$ with ``L'' and ``H'' denoting ``light'' and
``heavy''. $B_L^q$ and $B_H^q$ have different masses $M_{H,L}^q$ and
decay widths $\Gamma^q_{H,L}$.  $|M_{12}^q|\simeq \dm_q/2$ determines
the mass difference $\dm_q=M_H^q-M_L^q$, which coincides with the
well-measured \bbq\ oscillation frequency.  $\Gamma_{12}^q$ is much
smaller in magnitude than $M_{12}^q$ and enters the width difference
$\dg_q=\Gamma^q_L-\Gamma^q_H $ and the CP asymmetry in flavour-specific
decays, $a_{\rm fs}^q$ as
\begin{eqnarray}
\frac{\dg_q}{\dm_q} \; =\; - \real  \frac{\Gamma_{12}^q}{M_{12}^q} ,
  \qquad\qquad
  a^q_{\rm fs}
     \;=\;
    \imag \frac{\Gamma_{12}^q}{M_{12}^q}.
  \label{dgafs}
\end{eqnarray}
We know much less about $\Gamma_{12}^q$ than $M_{12}^q$: While $\dg_s$
is measured precisely,
\begin{eqnarray}
\dg_s &=& (0.0781 \pm 0.0035) \, \mbox{ps}^{-1}, \label{eq:exp}
\end{eqnarray}
experimental results on $\dg_d$, $a^d_{\rm fs}$, and $a^s_{\rm fs}$ are
all consistent with zero.  The 2025 HFLAV average \cite{HeavyFlavorAveragingGroupHFLAV:2024ctg} in
\eq{eq:exp} uses data from the D\O, CDF, ATLAS, LHCb, and CMS
experiments \cite{D0:2011ymu,CDF:2012nqr, ATLAS:2014nmm,
  LHCb:2014iah,CMS:2015asi,ATLAS:2016pno, LHCb:2016tuh, LHCb:2017hbp,
  LHCb:2019nin,ATLAS:2020lbz,CMS:2020efq,LHCb:2021wte,LHCb:2023sim,LHCb:2023xtc}.  On
the theory side, $\Gamma_{12}^q$ is much more difficult to calculate
than $M_{12}^q$: Firstly, the scale governing the perturbative QCD
corrections is $\mu\sim m_b$ in $\Gamma_{12}^q$, opposed to
$\mu\sim m_t$ in $M_{12}^q$. Secondly, unlike $M_{12}^q$, which only
involves a single dimension-6 operator, the Heavy Quark Expansion (HQE)
\cite{hqe,hqe2,hqe3,hqe4} of $\Gamma_{12}^q$ involves corrections of
order $2\lqcd/m_b={\cal O}(20\%)$ containing five matrix elements of
dimension-7 operators \cite{Beneke:1996gn}. The calculation of these
matrix elements with lattice QCD is difficult, because they mix with
dimension-6 operators requiring the control of effects which diverge
with one power of the inverse lattice spacing. Currently errors of order
40\% are quoted for these matrix elements \cite{Davies:2019gnp}.

For the discussion of $1/m_b$ corrections it is convenient to work with
$\Gamma_{21}^q$ rather than $\Gamma_{12}^{q}$.  We further specify to
$q=s$, for which the envisaged corrections are most urgently needed; the
generalization to $q=d$ is straightforward.  The $(\lqcd/m_b)^0$
leading-power term of the HQE for $\Gamma_{21}^s$ involves two
dimension-six $\Delta B=-2$ operators ($B$ denotes the beauty quantum
number)%
\bea%
\label{eq:defops}
Q&=&
           \bar{b}_i \gamma_{\mu} L s_i \, \bar{b}_j
           \gamma^{\mu} L s_j, \qquad\qquad %
\widetilde{Q}_S\;=\; \bar{b}_i L s_j \, \bar{b}_j L s_i %
\eea %%%
Here the $i,j$ are color indices and $L=(1 -\gamma_5)$.  The Wilson
coefficients of these operators have been calculated at next-to-leading
order (NLO) of QCD perturbation theory
\cite{Beneke:1998sy,Beneke:2003az,Ciuchini:2003ww,Lenz:2006hd,Gerlach:2021xtb,Gerlach:2022wgb}
and NNLO results have been first obtained in the limit of a large number
$N_f$ of flavors \cite{Asatrian:2017qaz} and subsequently for
contributions of the chromomagnetic operator
\cite{Asatrian:2017qaz,Gerlach:2022wgb}.  Finally the complete NNLO
contributions involving the dominant current-current operators have been
calculated \cite{Gerlach:2022hoj,Gerlach:2025tcx}.

The hadronic matrix elements of the operators in \eq{eq:defops}, which
are calculated with non-perturbative methods such as lattice QCD
\cite{Dowdall:2019bea} or QCD sum rules \cite{Kirk:2017juj}, are
commonly parametrized as%
\bea%%
\label{eq:defb}
\bra{B_s} Q (\mu) \ket{\ov B_s} = M^2_{B_s}\, f^2_{B_s}
\left(2+\frac{2}{N_c}\right) B(\mu), \quad \bra{B_s} \widetilde Q_S
(\mu)\ket{\ov B_s} = M^2_{B_s}\, f^2_{B_s} \left(1-\frac{2}{N_c}\right)
\widetilde B_S^\prime (\mu). %%
\eea %%
Here $M_{B_s}$ and $f_{B_s}$ are the mass and the decay constant of the
$B_s$ meson, $N_c$ is the number of colors and $\mu={\cal O}(m_b)$ is
the renormalization scale at which the matrix elements are
calculated. In lattice-gauge theory $\mu$ is the scale at which the
lattice continuum matching is performed.

The prediction of {Ref.\cite{Gerlach:2025tcx}} for $\dg_s$ reads 
\begin{eqnarray}
  \dg_s &=& \lt[ 0.077 \lt.\epm{0.005}{0.007}\rt|_{\rm scale}\lt.
              \pm 0.014  \rt|_{1/m_b} \pm \lt. 0.002 \rt|_{B,\widetilde
            B_S^\prime} \pm \lt.  0.001 \rt|_{\rm input} \rt]
            \mbox{ps}^{-1}.
            \label{eq:dgnnlo}
\end{eqnarray}
Here the first uncertainty shows the dependence on the renormalization
scale of the NNLO result for the leading-power term in
$\Gamma_{12}^s$. The second uncertainty stems from the $1/m_b$ term. The
third and last uncertainties refer to the hadronic matrix elements in
\eq{eq:defb} and the parametric input, namely quark masses and CKM
elements, respectively.  The quoted number $0.014\,\mbox{ps}^{-1}$ reflects the
uncertainties from both the lattice calculation \cite{Davies:2019gnp}
and the perturbative LO calculation of the coefficients
\cite{Beneke:1996gn}. In fact both uncertainties cannot be separated,
because a meaningful lattice-continuum matching requires coefficients
calculated at least to NLO.

The large uncertainty in \eq{eq:dgnnlo} calls for an NLO
calculation of the coefficients of the dimension-7 operators occurring in
the $1/m_b$ corrections. In this paper we present a first step in that
direction by calculating the mixing of these operators with the
dimension-6 operators $Q$ and $\widetilde Q_S$ and the associated
renormalization prescription. The renormalization constants presented
below are needed  for both the lattice-continuum matching and the
calculation of the NLO coefficients. Our paper is organized as follows:
In Sec.~\ref{pr} we list the operators and summarize preliminary work.
Secs.~\ref{rn} and \ref{fc} are devoted to the derivation of the
renormalization constants and the calculation of matrix elements in the
limit of a large number of colors, $N_c\to \infty$, respectively.  In Sec.~\ref{cn} we
conclude.

\section{Preliminaries\label{pr}}
$\Gamma_{21}^s=\Gamma_{12}^{s*}$ is decomposed as \cite{Beneke:1998sy}
\begin{eqnarray}
  \Gamma_{21}^s &=& - (V_{cb}^*V_{cs})^2\Gamma^{cc}_{21} 
                  - 2  V_{cb}^*V_{cs} V_{ub}^*V_{us} \Gamma_{21}^{uc} 
                  - (V_{ub}^*V_{us})^2\Gamma^{uu}_{21} 
                  \,.
                    \label{eq::Gam12}
\end{eqnarray}
The individual terms read 
\begin{eqnarray}
  \Gamma_{21}^{ab} 
  &=& \frac{G_F^2m_b^2}{24\pi M_{B_s}} \left[ 
      H^{ab}(z)   \langle \bar{B}_s|Q| B_s\rangle
      + \widetilde{H}^{ab}_S(z)  \langle \bar{B}_s|\widetilde{Q}_S|B_s \rangle
      \right]
      + \widetilde\Gamma_{21,1/m_b}^{ab} + \ldots \,
      \label{eq::Gam^ab}
\end{eqnarray}
Here the first term is the leading-power term with various
contributions to the coefficients $ H^{ab}$ and $ \widetilde{H}^{ab}_S$
calculated to NLO and NNLO as described in the
Introduction. $z=m_c^2/m_b^2$ encodes the heavy-quark mass ratio.
$ \widetilde\Gamma_{21,1/m_b}^{ab}$ contains the $1/m_b$ corrections. We
use the definition of Ref.~\cite{Lenz:2006hd} which differs from the
original one in Ref.~\cite{Beneke:1996gn}, because the original
calculation was not done in the basis $(Q,\widetilde Q_S)$ for the
dimension-6 operators used since 2006. The dots denote terms of order
$1/m_b^2$ and higher. One has
\begin{eqnarray}
  \widetilde{\Gamma}_{12,1/m_b}^{ab} 
&=&  \frac{G_F^2 m_b^2}{24 \pi M_{B_s}} 
   \lt[ g_0^{ab} (z) \bra{{\bar B}_s} R_0 \ket{{B}_s}  \, + \,
   \sum_{j=1}^3 \lt[ g_j^{ab} (z) \bra{{\bar B}_s} R_j \ket{{B}_s} + 
       \widetilde{g}_j^{ab} (z) \bra{{\bar B}_s} \widetilde{R}_j \ket{{B}_s} 
       \rt]
      \rt] \label{ga12m}
\end{eqnarray}
the coefficients $g_j^{ab}(z)$, derived from the results in
Refs.~\cite{Beneke:1996gn} and \cite{Dighe:2001gc}, are known to LO in
$\alpha_s$ and can be found in Eqs.~(25) and (26) of Ref.~\cite{Lenz:2006hd}. 

The dimension-7 operators entering \eq{ga12m} are \cite{Beneke:1996gn}%
\bea%
R_0 &=& Q_S+\tilde{Q}_S+\frac{1}{2}Q \qquad\qquad\mbox{with }\quad
% Q_S=(\bar{b}_is_i)_{S-P}\;(\bar{b}_j s_j)_{S-P},  \\
Q_S\;=\; \bar{b}_i L s_i \,  \bar{b}_j L s_j,  \label{eq:defr0} \\
R_1 &=& \frac{m_s}{m_b} \,
  \bar{b}_i L s_i \,  \bar{b}_i R s_i \label{eq:defr1}\\
R_2 &=& \frac{1}{m_b^2} \left(\bar{b}\overleftarrow{D}_{\rho}\right)_i
  \gamma^{\mu} L \left(D^{\rho} s\right)_i \, \bar{b}_j
  \gamma_{\mu} L s_j,
  \\
R_3 &=& \frac{1}{m_b^2}
  \left(\bar{b}\overleftarrow{D}_{\rho}\right)_i L \left(D^{\rho}
    s\right)_i \, \bar{b}_j L s_j, %
 \eea%
 with $L,R=(1\mp \gamma_5)$ and $D_{\rho} = \partial_{\rho}-ig_sA_{\rho}^aT^a$.
 There are further color-rearranged operators $\widetilde{R}_i$ for
 $i=1,2,3$, obtained from the expressions above  by
 interchanging $s_i$ and $s_j$. At order $1/m_b$ only a subset of these
 operators is needed, as some of their linear combinations have $1/m_b^2$
 suppressed matrix elements; in particular one has $\langle R_2\rangle =
 -\langle \tilde R_2 \rangle + {\cal O}  (1/m_b^2)$. 
 
 In the power counting factors of $m_s$ and derivatives $\partial^\mu$
 on the $s$ field count as $\lqcd$ as they lead to factors of $m_s$ or
 $p_s^\mu$ in the tree-level matrix elements between quark states,
 accompanied by factors of $1/m_b$ in the coefficients. The bare loop
 matrix elements do not have this feature, because $\partial^\mu s$ can
 give a power of the loop momentum with the subsequent loop integration
 giving a power of $m_b$ instead of $m_s$ or $p_s^\mu$. Clearly, this is
 an ultraviolet (UV) effect stemming from the region of large loop
 momenta. We therefore {expect to be able to} absorb these terms into
 finite {counterterms} to restore the correct power counting of the
 loop-corrected matrix elements.  To this end we employ dimensional
 regularization in $D=4-2\epsilon$ dimensions and define
\begin{eqnarray}
  R_j^{\mathrm{ren}}
  &=& Z_{jQ} Q^{\mathrm{bare}} + Z_{j\widetilde Q_S} \widetilde
      Q_S^{\mathrm{bare}}
      + \sum_{k=0}^3 \lt[     Z_{jk} R_k^{\mathrm{bare}} +
        Z_{jk+3} \widetilde R_k^{\mathrm{bare}} \rt],  \label{eq:rb}\\
 \widetilde R_j^{\mathrm{ren}}
  &=&  \widetilde Z_{jQ} Q^{\mathrm{bare}} + \widetilde Z_{j\widetilde Q_S} \widetilde
      Q_S^{\mathrm{bare}}
      + \sum_{k=0}^3 \lt[   \widetilde   Z_{jk} R_k^{\mathrm{bare}} +
      \widetilde Z_{jk+3} \widetilde R_k^{\mathrm{bare}} \rt] ,
  \label{eq:rtb}
\end{eqnarray}
and
\begin{eqnarray}
 Z_{jQ} &=&  \frac{\alpha_s}{4\pi} \lt[ \frac{Z_{jQ}^{(1 1)}}{\epsilon} +
            Z_{jQ}^{(10)} \rt]  + {\cal O} (\alpha_s^2), \qquad\qquad
  Z_{j\widetilde Q_S} \;=\;  \frac{\alpha_s}{4\pi} \lt[ \frac{
                       Z_{j\widetilde Q_S}^{(1 1)}}{\epsilon} +
            Z_{j \widetilde Q_S}^{(10)} \rt] + {\cal O} (\alpha_s^2) \label{eq:zexp}
\end{eqnarray}
with analogous definitions of $\widetilde Z_{jQ}$ and
$\widetilde Z_{j\widetilde Q_S}$.  We replace $\mu^\epsilon g$ by
$(\mu e^{-\gamma_E}/(4\pi)) ^\epsilon g$ in the Feynman rules for the
gauge coupling to remove the Euler constant $\gamma_E$ and the term
$\log(4\pi)$ from our matrix elements.  $\mu$ is the renormalization
scale at which the matrix elements are calculated and
$\alpha_s=g^2/(4\pi)$ is to be read as $\alpha_s(\mu)$ everywhere.  We
further write
\begin{eqnarray}
  \langle R_j \rangle^{\mathrm{ren}}
   &=& \langle R_j \rangle^{(0)} + \frac{\alpha_s}{4\pi} \langle R_j^{\mathrm{ren}}
       \rangle^{(1)} + {\cal O}(\alpha_s^2) 
\end{eqnarray}
and similarly for $ \langle \widetilde R_j \rangle^{\mathrm{ren}}$ and
the bare matrix elements. We will expand the one-loop matrix elements in
terms of tree-level matrix elements counting
$\langle R_j\rangle^{(0)} ={\cal O}(1/m_b)$ and
$\langle Q \rangle^{(0)}$,
$\langle \widetilde Q_S \rangle^{(0)}={\cal O}(m_b^0)$.

In this paper only terms of $\langle R_j \rangle^{(1)}$ proportional to
$\langle Q \rangle^{(0)}$ and $\langle \widetilde Q_S \rangle^{(0)}$ are
to be considered.  In the $\ov{\rm MS}$ scheme the finite pieces in
\eq{eq:zexp} all vanish but $\langle R_j \rangle^{(1)}$ contains these
unwanted matrix elements of dimension-6 operators.  The purpose of
this paper is the determination of $ Z_{jQ}^{(10)}$, $Z_{j\widetilde Q_S}^{(10)}$,
$\widetilde Z_{jQ}^{(10)}$,  and $\widetilde Z_{j\widetilde Q_S}^{(10)}$
% such that
in
\begin{eqnarray}
  \delta R_j &\equiv& \frac{\alpha_s }{4\pi} \delta R_j^{(1)}, \qquad
  \; \delta R_j^{(1)} \equiv 
                      \lt[ \frac{Z_{jQ}^{(11)}}{\epsilon} +
      Z_{jQ}^{^(10)} \rt] Q  +
     \lt[ \frac{Z_{j\widetilde Q_S}^{(11)}}{\epsilon} +
      Z_{j\widetilde Q_S}^{(10)} \rt] \widetilde Q_S 
  \\
  \delta \widetilde R_j &\equiv& \frac{\alpha_s}{4\pi} \delta \widetilde
                                 R_j^{(1)}, \qquad
    \delta \widetilde R_j^{(1)} \; \equiv   \lt[ \frac{\widetilde Z_{jQ}^{(11)}}{\epsilon} +
      \widetilde Z_{jQ}^{(10)} \rt] Q +
     \lt[ \frac{\widetilde Z_{j\widetilde Q_S}^{(11)}}{\epsilon} +
      \widetilde Z_{j\widetilde Q_S}^{(10)} \rt]  \widetilde
      Q_S
   \label{eq:rren}    
\end{eqnarray}
such that
\begin{eqnarray}
  \langle R_j^{\mathrm{ren}} \rangle^{(1)} &=&  \langle R_j^{\mathrm{bare}}     \rangle^{(1)} +  \langle
      \delta R_j         \rangle^{(1)} \;=\;  {\cal O} \lt(
                                               \frac1{m_b}\rt) \label{eq:fct}
\end{eqnarray}
and analogously for $\widetilde R_j^{\mathrm{ren}}$.

For $R_0$ this task has already been completed  in
Ref.~\cite{Beneke:1998sy}. For the dimension-6 basis $(Q,\widetilde
Q_S)$ the counterterm reads  
\begin{eqnarray}
  \delta R_0^{(1)} 
  &=&
      \lt( (N_c+1) \lt(       \frac{1}{2\epsilon} + \log \frac{\mu}{m_b}\rt)
      + \frac{N_c^2+2 N_c -2}{N_c} \rt) Q \no\\
  &&  \,+\,
      2 (N_c+1)
      \lt( \frac{1}{\epsilon}+ 2 \log \frac{\mu}{m_b} + 1 \rt) \widetilde Q_S. 
\end{eqnarray}
$C_F=(N_c^2-1)/(2N_c)$ is a color factor. The result for $N_c=3$ was
given in Ref.~\cite{Lenz:2006hd}.
$ \delta R_0 $ has been
calculated to order $\alpha_s^2N_f$ in Ref.~\cite{Asatrian:2017qaz};
the full result to order $\alpha_s^2$ is presented in
Ref.~\cite{Gerlach:2025tcx}. 

$\delta R_1^{(1)}=\delta \widetilde R_1^{(1)}=0$, because the factor of
$m_s$ in the definition of these operators (see {\eq{eq:defr1}})
prevents any mixings with $Q$ or $\widetilde Q_S$.

% \section{\bf Renormalization of operators $R_2$, $\widetilde{R}_2$ and
%   $R_3$, $\widetilde{R}_3$ }
\section{Renormalization of operators $R_2$, $\widetilde{R}_2$, $R_3$,  and $\widetilde{R}_3$\label{rn}}

In this section we calculate $\delta R_{2,3}^{(1)}$ and  $\delta
\widetilde R_{2,3}^{(1)}$. 
\begin{figure}[t]
\vspace{-5.2cm}
\hspace{-2.7cm}
\includegraphics[scale=1.0]{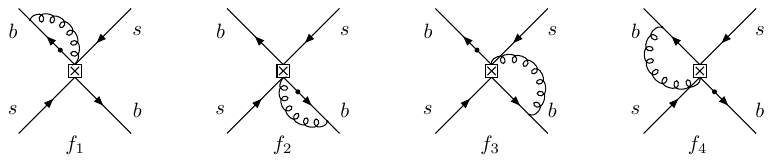}
\vspace{-21.8cm}
\caption{\label{fig1} Infrared-finite diagrams % (1st and 2nd Wick
  % contractions)
  originating from the couplings of the operators $\widetilde{R}_2$
  and $\widetilde{R}_3$ with a gluon field and a derivative acting on a $b$ field.
  The $b$ line with  derivative is marked with  a dot.}
~\\[-5mm]
\hrule
\end{figure}
We encounter two types of contributions to the desired finite
renormalization.  \fig{fig1} shows those Feynman diagrams which
have a single derivative acting only on a $b$ field and a gluon field
from $D^\rho s$.  
\fig{fig2}
depicts the diagrams which involve both $\partial_\rho b$ and $\partial^\rho s$.
In the integrals corresponding to these Feynman diagrams, we set
the small strange quark momentum $p_s$ and the strange quark mass $m_s$,
which we count as $\Lambda_{QCD}$, equal to zero. Terms proportional to
 $p_s$ or $m_s$ will only be needed for the calculation of
 $Z_{jk}$,  $\widetilde Z_{jk}$ in \eqsand{eq:rb}{eq:rtb}, but not for
 the mixings into $Q$, $\widetilde Q_S$. Now $\partial^\rho s$ can only
 give a non-zero contribution when the derivative gives a power of the
 loop momentum, reducing the number of diagrams to the four  of \fig{fig1}. 
\begin{figure}[t]
\vspace{-2.1cm}
\hspace{-2.7cm}
\includegraphics[scale=1.0]{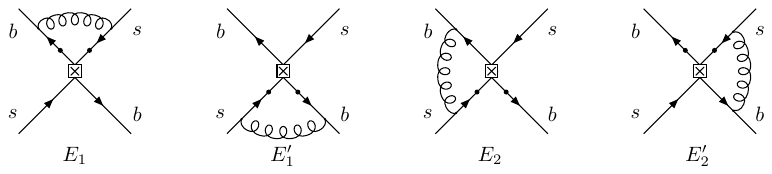}
\vspace{-24.8cm}
\caption{\label{fig2} Diagrams $E_1$, $E_1'$, $E_2$ and $E_2'$ which contribute to the finite renormalization of operators $R_2$, $R_3$, $\widetilde{R}_2$ and $\widetilde{R}_3$. The derivatives, acting on the $b$ and $s$ fields, are denoted as dots on the corresponding lines.}
~\\[-5mm]
\hrule

\end{figure}

\subsection{Evanescent operators\label{sec:eva}}
Our results depend on the chosen renormalization scheme, characterized by
the use of an anticommuting $\gamma_5$ (NDR scheme) and the
following definitions of the evanescent operators
\cite{Buras:1989xd,Beneke:1998sy}
\begin{eqnarray}
  E_1[Q] &\equiv&\;            \bar{b}_i \gamma_{\mu} \gamma_{\nu}
                              \gamma_{\rho} L s_i \, \bar{b}_j
             \gamma^{\rho} \gamma^{\nu}\gamma^{\mu} L s_j
             \, -\,  (4 -8 \epsilon) Q \; + \; \mathcal{O} (\epsilon^2), \label{eq:evq} \\
E_2[Q_S] & \equiv &\; \bar{b}_i \gamma_{\mu} \gamma_{\nu}L
                           s_i
                           \, \bar{b}_j \gamma_{\nu} \gamma_{\mu} L s_j
                         \,+\, (8-8\epsilon) \widetilde Q_S  \;+\; \mathcal{O}
               (\epsilon^2),
               \label{eq:evqs} \\
E_2[\widetilde{Q}_S] & \equiv &\; \bar{b}_i \gamma_{\mu} \gamma_{\nu}L
                           s_j
                           \, \bar{b}_j \gamma_{\nu} \gamma_{\mu} L s_i 
                         \,+\, (8-8\epsilon) Q_S \;+\; \mathcal{O} (\epsilon^2) .\label{eq:evqst} 
\end{eqnarray}
Evanescent operators receive finite counterterm such that their
renormalized matrix elements vanish for $\epsilon=0$. The coefficients
multiplying $\epsilon$ in the definitions in \eqsto{eq:evq}{eq:evqst},
however, combine with $1/\epsilon$ poles to finite contributions
affecting our $Z_{jQ}^{(10)},\ldots\widetilde Z_{j\widetilde
  Q_S}^{(10)}$. In the practical calculation, for instance,
\eqsand{eq:evqs}{eq:evqst} mean that one replaces  structures with
two Dirac matrices on each fermion line as follows:
\begin{eqnarray}
  \lt[ \gamma_{\mu} \gamma_{\nu}L\rt]_{lm}  &\otimes&
  \lt[\gamma_{\nu} \gamma_{\mu}L\rt]_{np} 
  \to  
  (8- 8\epsilon) L_{lp} \otimes L_{nm}\,, \no \\
         \lt[ \gamma_{\mu} \gamma_{\nu}L\rt]_{lm}  &\otimes&
     \lt[\gamma_{\mu} \gamma_{\nu}L\rt]_{np}
     \to
     -  (8- 8\epsilon) L_{lp} \otimes L_{nm} + (8-4\epsilon) L_{lm} \otimes L_{np}
\label{eq:exev}
\end{eqnarray}  
with the second prescription obtained from the first one by a simple
anti-commutation of the two Dirac matrices. $l,\ldots p$ are Dirac
indices.  Note that these relations involve Fierz transformations
(interchange of {indices} $m$ and $p$ on the fermion lines), which a
    priori do not hold in $D\neq 4$ dimensions. The choice of the
    coefficients of $\epsilon$ in \eqsand{eq:evqs}{eq:evqst} insures
    that the renormalized one-loop matrix elements obey Fierz symmetry,
    \textit{i.e.}\ these matrix elements are equal to those of the
    operators $Q^F=\widetilde Q$, $Q_S^F$, $\widetilde Q_S^F$, found
    from $Q,Q_S,\widetilde Q_S$ by Fierz transformations.

We define the evanescent operators for the dimension-7 operators as%
\bea E_1[R_2] &\equiv&\; \frac{1}{m_b^2}
\left(\bar{b}\overleftarrow{D}_{\rho}\right)_i \gamma_{\mu} \gamma_{\nu}
\gamma_{\alpha} L \left(D^{\rho} s\right)_i \, \bar{b}_j \gamma^{\alpha}
\gamma^{\nu}\gamma^{\mu} L s_j \, -\, (4 -a_2 \epsilon) R_2 \; + \;
\mathcal{O} (\epsilon^2),
\\
E_2[R_2] &\equiv&\; \frac{1}{m_b^2}
\left(\bar{b}\overleftarrow{D}_{\rho}\right)_i \gamma_{\mu} \gamma_{\nu}
L \left(D^{\rho} s\right)_i \, \bar{b}_j \gamma^{\nu}\gamma^{\mu} L s_j
\, + \, (8 - b_2 \epsilon) \widetilde{R}_3 \; + \; \mathcal{O}
(\epsilon^2),
\\
E_2[R_3] &\equiv&\; \frac{1}{m_b^2}
\left(\bar{b}\overleftarrow{D}_{\rho}\right)_i \gamma_{\mu} \gamma_{\nu}
L \left(D^{\rho} s\right)_i \, \bar{b}_j \gamma^{\nu}\gamma^{\mu} L s_j
\, + \, (8 - b_3 \epsilon) \widetilde{R}_3 \; + \; \mathcal{O}
(\epsilon^2), \label{eq:defevr}%
\eea%
and similar for color-rearranged operators $\widetilde{R}_n$
($n = 2,3$), obtained from the expressions above by interchanging $s_i$
and $s_j$, and replacing constants $a_n$ and $b_n$ by $\widetilde{a}_n$
and $\widetilde{b}_n$ correspondingly. The numerical values of these
constants are to be fixed later. {If we impose the requirement of Fierz
  invariance, in analogy to $Q$ and $\widetilde Q_S$ above, on $R_j$ and
  $\widetilde{R}_j$, we} obtain $a_2=\widetilde{a}_2=12$ and
$b_2=\widetilde{b}_2=4$, while the Fierz invariance of operators $R_3$
and $\widetilde{R}_3$ yields $b_3=\widetilde{b}_3=10$. Thus, since
  $b_2=\widetilde b_2$ and $b_3=\widetilde b_3$ differ from each other
  and from the corresponding coefficient in \eq{eq:evqst}, the
  replacement rule in \eq{eq:exev} is different for {$Q_S$},
  $\widetilde Q_S$, ${R_2}$, $\widetilde R_2$, ${R_3}$, and
  $\widetilde R_3$, if one applies the criterion that
  $\ov{\rm MS}$-renormalized matrix elements of these operators should
  be equal to those of the operators obtained from them by Fierz
  transformations. {In Sec.~\ref{fc} we will see that Fierz symmetry
  matters for correct results for $\widetilde Z_{jQ}^{(10)}$ and
  $\widetilde Z_{j\widetilde Q_S}^{(10)}$.}

\subsection{Results for the counterterms}
Collecting all the results, which contribute to the finite
renormalization of the operator $R_2$ in the $\overline{\rm MS}$ scheme
we obtain%
\bea%
\nonumber \delta\langle R_2^{\ov{\mathrm{MS}}} \rangle^{(1)} &=&
\phantom{+ } \left(\left(2-\frac{2}{N_c}-\frac{8 C_F}{3}\right) \log
  \frac{\mu}{m_b} + 2 - \frac{a_2}{24} - \left(2 -
    \frac{a_2}{24}\right) \frac{1}{N_c}-\frac{38 C_F}{9}\right)
\langle Q\rangle^{(0)}
\\
\nonumber && + \left(\left(\frac{2}{3}-\frac{4}{3 N_c}\right) \log
  \frac{\mu}{m_b} + \frac{2}{9} -
  \left(\frac{4}{9}-\frac{b_2}{12}\right) \frac{1}{N_c}\right)
\langle\widetilde{Q}_S\rangle^{(0)}
\\
&& + \left(\left(\frac{4}{3}-\frac{2}{3 N_c}+\frac{4 C_F}{3}\right) \log
  \frac{\mu}{m_b} + \left(\frac{4}{9}-\frac{b_2}{12}\right) -
  \frac{2}{9 N_c}-\frac{2 C_F}{9}\right) \langle Q_S\rangle^{(0)}
\\
&=& \phantom{+ } \left(\left(\frac{4}{3}-\frac{5}{3 N_c}-\frac{10
      C_F}{3}\right) \log \frac{\mu}{m_b} +\frac{16}{9} +
  \frac{b_2-a_2}{24} -
  \left(\frac{17}{9}-\frac{a_2}{24}\right) \frac{1}{N_c}-\frac{37
    C_F}{9} \right) \langle Q\rangle^{(0)}
\no \\
&& + \left(\left(-\frac{2}{3}-\frac{2}{3 N_c}-\frac{4 C_F}{3}\right)
  \log \frac{\mu}{m_b} - \frac{2}{9} + \frac{b_2}{12} -
  \left(\frac{2}{9} - \frac{b_2}{12} \right) \frac{1}{N_c}+
  \frac{2 C_F}{9}\right) \langle \widetilde
Q_S\rangle^{(0)}. \label{eq:r2ms} \eea%
In \eq{eq:r2ms} we have used \eq{eq:defr0} to eliminate
$Q_S=R_0- \widetilde Q_S -Q/2 $ and used that
$\langle R_0\rangle^{(0)}=0$ for $p_s=0$. Thus%
\bea %
\label{R2renres}
\nonumber \delta R_2^{(1)} &=& \phantom{+ } \left(\left(-\frac{4}{3}+\frac{5}{3 N_c}+\frac{10 C_F}{3}\right) \left(\frac{1}{2\epsilon} + \log \frac{\mu}{m_b}\right)  - \frac{16}{9} - \frac{b_2-a_2}{24} + \left(\frac{17}{9}-\frac{a_2}{24}\right) \frac{1}{N_c} + \frac{37 C_F}{9}
\right)  Q
\\
&& + \left(\left(\frac{2}{3}+\frac{2}{3 N_c}+\frac{4 C_F}{3}\right)
 \left(\frac{1}{2\epsilon} + \log \frac{\mu}{m_b}\right)  + \frac{2}{9} - \frac{b_2}{12} + \left(\frac{2}{9} - \frac{b_2}{12} \right) \frac{1}{N_c} - \frac{2 C_F}{9}\right) \widetilde{Q}_S %
\eea%
and one arrives at
$\langle R_2^{\mathrm{bare}}+\delta R_2\rangle^{(1)}={\cal O} (1/m_b)$
as desired.

Similarly for operator $R_3$ we obtain%
\bea%
\no \langle R_3^{\ov{\mathrm{MS}}} \rangle^{(1)} &=& \phantom{+ }
\left(\left(\frac{1}{6}-\frac{1}{6 N_c}\right) \log \frac{\mu}{m_b} + \frac{1}{18} - \frac{1}{18N_c} \right) \langle Q\rangle^{(0)}
\\
&& \nonumber + \left(\left(\frac{4}{3}+\frac{2}{3 N_c}\right) \log \frac{\mu}{m_b} + \frac{29}{18} + \left(\frac{5}{9} - \frac{b_3}{24}\right) \frac{1}{N_c} \right) \langle\widetilde{Q}_S\rangle^{(0)}
\\
&& + \left( -\left(\frac{2}{3}+\frac{4}{3 N_c}\right) \log \frac{\mu}{m_b} - \frac{5}{9} + \frac{b_3}{24} - \frac{29}{18N_c} - 3 C_F \right) \langle Q_S\rangle^{(0)},
 \label{eq:r3ms}%
\eea%
entailing %
\bea%
\no \delta R_3^{(1)} &=& - \left( \left(\frac{1}{2}+\frac{1}{2 N_c}\right) \left(\frac{1}{2\epsilon} + \log \frac{\mu}{m_b}\right) + \frac{1}{3} -\frac{b_3}{48} + \frac{3}{4N_c} +\frac{3C_F}{2} \right) Q
\\
&&  -\left(\left(2+\frac{2}{N_c}\right) \left(\frac{1}{2\epsilon} + \log \frac{\mu}{m_b}\right) + \frac{13}{6} -\frac{b_3}{24} + \left(\frac{13}{6} -\frac{b_3}{24}\right) \frac{1}{N_c} + 3 C_F \right) \widetilde{Q}_S.
\label{R3renres} % 
\eea%
For  $\widetilde{R}_2$ we find% 
\bea %
\no \langle \widetilde R_2^{\ov{\mathrm{MS}}} \rangle^{(1)} 
&=&  \phantom{+ }
  \left(\left(-\frac{4}{3}+\frac{4}{3 N_c}+4 C_F\right) \log \frac{\mu}{m_b} - \frac{19}{9} + \frac{19}{9 N_c} + \left(4-\frac{\widetilde{a}_2}{12}\right) C_F\right) \langle Q\rangle^{(0)}
\\
&& \nonumber + \left(\left(-\frac{2}{3 N_c}+\frac{4 C_F}{3}\right) \log \frac{\mu}{m_b} + \frac{1}{9 N_c}+\frac{4 C_F}{9}\right) \langle\widetilde{Q}_S\rangle^{(0)}
\\
&& + \left(\left(\frac{2}{3}+\frac{8 C_F}{3}\right) \log \frac{\mu}{m_b} - \frac{1}{9} + \left(\frac{8}{9}-\frac{\widetilde{b}_2}{6}\right)C_F\right) \langle Q_S\rangle^{(0)}
, \label{eq:r2tms} %
\eea%
so that%
\bea%
\no \delta \widetilde{R}_2^{(1)} &=& \phantom{+ } \left(\left(\frac{5}{3}-\frac{4}{3 N_c}-\frac{8 C_F}{3}\right) \left(\frac{1}{2\epsilon} + \log \frac{\mu}{m_b}\right) + \frac{37}{18} - \frac{19}{9 N_c} - \left(\frac{32}{9} +\frac{\widetilde{b}_2-\widetilde{a}_2}{12} \right)C_F \right) Q
\\
&& + \left(\left(\frac{2}{3}+\frac{2}{3 N_c}+\frac{4 C_F}{3}\right) \left(\frac{1}{2\epsilon} + \log \frac{\mu}{m_b} \right) -\frac{1}{9}-\frac{1}{9 N_c} + \left(\frac{4}{9} - \frac{\widetilde{b}_2}{6}\right) C_F \right) \widetilde{Q}_S.
\label{R2Crenres} %
\eea%
Finally for operator $\widetilde{R}_3$ we obtain %
\bea%
\label{R3tMS}
\no 
\langle \widetilde{R}_3^{\ov{\mathrm{MS}}} \rangle^{(1)}
 &=&\phantom{+ } \left(\frac{1}{3} C_F \log \frac{\mu}{m_b} + \frac{C_F}{9}\right) \langle Q\rangle^{(0)}
 + \left(\frac{8 C_F}{3} \log \frac{\mu}{m_b} + \frac{3}{2N_c}+\frac{29 C_F}{9}\right) \langle\widetilde{Q}\rangle^{(0)}_S
\\
&& + \left(-\frac{4 C_F}{3} \log \frac{\mu}{m_b} - \frac{3}{2} - \left(\frac{10}{9} -\frac{\widetilde{b}_3}{12} \right)C_F\right) \langle Q_S\rangle^{(0)}, \label{R3Crenresmsbar}  %
\eea
leading to the counterterm%
\bea%
\no
\delta  \widetilde{R}_3^{(1)}
&=& -\left( C_F \left(\frac{1}{2\epsilon} + \log \frac{\mu}{m_b}\right) + \frac{3}{4} + \left(\frac{2}{3}-\frac{\widetilde{b}_3}{24}\right) C_F \right) Q
\\
&& - \left(4C_F \left(\frac{1}{2\epsilon} + \log \frac{\mu}{m_b}\right) + \frac{3}{2} + \frac{3}{2N_c} + \left(\frac{13}{3}-\frac{\widetilde{b}_3}{12}\right) C_F \right) \widetilde{Q}_S.
\label{R3Crenres}
\eea%
The results for the renormalization constants can be simply read off
from Eqs.~(\ref{R2renres}), (\ref{R3renres}), (\ref{R2Crenres}), and
(\ref{R3Crenres}); we summarize 
$Z_{jQ}^{(10)}$, $Z_{j\widetilde Q_S}^{(10)}$, 
$\widetilde Z_{jQ}^{(10)}$, and $\widetilde Z_{j\widetilde Q_S}^{(10)}$
in \tab{tab}.

\section{Matrix elements for $\mathbf{N_c\to \infty}$ \label{fc}}
In the limit $N_c\to\infty$ one can calculate the hadronic matrix
elements exactly, because they factorize into the product of two current
operators which can be expressed in term of the $B_s$ meson decay
constant $f_{B_s}$. {In this section we do this calculation for the
  $\ov{\rm MS}$-renormalized one-loop matrix elements of the studied
  dimension-7 operators. These calculations provide sophisticated checks
  of the conceptual and calculational correctness of our results and
  shed light on the issue of allowed choices for the parameters
  $\widetilde{a}_2$ and $\widetilde{b}_{2,3}$ in the definitions of the
  evanescent operators in \eq{eq:defevr}. The latter feature has to do
  with the fact that the large-$N_c$ calculation of the matrix elements
  of the colour-flipped operators $\widetilde{R}_{2,3}$ involves a
  four-dimensional Fierz transform.}

In Ref.~\cite{Beneke:1998sy} the limit $N_c\to\infty$ has lead to a
non-trivial result of the calculation of $\delta R_0$: In the
$\ov{\rm MS}$ scheme the factorized matrix element
$\langle \bar B_s | R_0^{{\ov{\rm MS}}} | B_s\rangle^{\rm fac}$ is proportional to
$(\bar m_b +\bar m_s)^2-M_{B_s}^2$ with the $\ov{\rm MS}$ quark masses
$\bar m_{b,s}$. This factor does not scale like $m_b \lqcd$ as
required for the proper power counting, one needs instead a (properly
renormalon-subtracted) pole mass $m_b^{\rm pole}$. The contribution from
$\delta R_0$ was found to fix this problem by providing the $\alpha_s$
correction which changes $\bar m_b^2-M_{B_s}^2$ to $m_b^{\rm pole\,2}
-M_{B_s}^2$. So the large-$N_c$ check illustrates that
$\langle \bar B_s | R_0^{\ov{\rm MS}} | B_s\rangle$ does not scale like
$\lqcd/m_b$, while
$\langle \bar B_s | R_0^{\rm ren} | B_s\rangle=\langle \bar B_s |
R_0^{\ov{\rm MS}} +\delta R_0 | B_s\rangle$ does.

To calculate the factorized matrix elements one uses
\begin{eqnarray}
\label{matelR2}
&& \langle 0 | \bar{b} \gamma_{\mu} \gamma_5 s (x) | B_s \rangle = if_{B_s} P_{\mu} e^{-iPx},
\qquad\qquad~\,
\langle \bar{B}_s | \bar{b} \gamma_{\mu} \gamma_5 s (x) | 0 \rangle = -if_{B_s} P_{\mu} e^{iPx},
\\
\label{matelS}
&& \langle 0 | \bar{b} \gamma_5 s (x) | B_s \rangle = -if_{B_s} \frac{M_{B_s}^2}{m_b+m_s} e^{-iPx},
\qquad
 \langle \bar{B}_s | \bar{b} \gamma_5 s (x) | 0 \rangle = -if_{B_s} \frac{M_{B_s}^2}{m_b+m_s} e^{iPx},
\end{eqnarray}
where {$\ket{B_s}=\ket{B_s(P)}$, $\ket{\bar B_s}=\ket{\bar B_s(P)}$}
    and the results in \eq{matelS} are obtained by applying
    $\partial^\mu$ to \eq{matelR2} and using
    $\partial^\mu \bar{b} \gamma_{\mu} \gamma_5 s = \bar b
    (\overleftarrow{D}^{\mu} + D^\mu) \gamma_{\mu} \gamma_5 s$.  With
    these results one finds the factorized matrix elements
    \cite{Beneke:1996gn,Lenz:2006hd}
\begin{eqnarray}
  \langle \bar{B}_s | Q | B_s \rangle^{\mathrm{fac}}
  &=& f_{B_s}^2 M_{B_s}^2\lt( 2+ \frac{2}{N_c} \rt) , \quad
      \langle \bar{B}_s | \widetilde Q_S^{\ov{\rm MS}} (\mu) | B_s \rangle^{\mathrm{fac}}
      \; =\;  f_{B_s}^2 M_{B_s}^2 \frac{M_{B_s}^2}{(\ov m_b+\ov
      m_s)^2}\lt( 1- \frac2{N_c} \rt). \quad
      \label{eq:facme}
\end{eqnarray}

\subsection{Operator $R_2$}
In the large-$N_c$ limit we obtain from \eq{eq:r2ms} with \eq{eq:facme}:
\bea% 
\label{resR2CF}
\langle \bar B_s| R_2^{\ov{\rm MS}} |B_s\rangle^{\mathrm{large}\, N_c}
& = &
  f_{B_s}^2M_{B_s}^2 \frac{\alpha_s\, N_c}{4\pi} \cdot \left[ 4 \log
 \frac{m_b}{\mu} -4 \right]  \,+ \, \mathcal{O} (\alpha_s^2,1/m_b) .% 
\eea%
The  large-$N_c$ matrix element of $m_b^2 R_2$ reads
\begin{eqnarray}
\label{factR2}
  m_b^2 \langle \bar B_s | R_2| B_s \rangle^{\mathrm{large}\,N_c}
  &=&  \langle \bar{B}_s | \bar{b} \gamma_{\mu} \gamma_5 s | 0 \rangle
      \langle 0 | \bar{b} \overleftarrow{D}_{\rho} \gamma_{\mu} \gamma_5
       D^{\rho} s | B_s \rangle +
     \langle \bar B_s  | \bar{b} \overleftarrow{D}_{\rho} \gamma_{\mu} \gamma_5
       D^{\rho} s | 0 \rangle \langle 0 | \bar{b} \gamma_{\mu} \gamma_5 s | B_s \rangle. \quad
\end{eqnarray}
In order to calculate the current matrix element with derivatives, one
applies $\partial_{\rho}\partial^{\rho}$ to \eq{matelR2} giving the
desired term twice and terms with two covariant derivatives acting on either
$\bar b$ or $s$:
\begin{eqnarray}
\label{factorR2}
&& 2 \langle 0 | \bar{b} \overleftarrow{D}_{\rho} \gamma_{\mu} \gamma_5 D^{\rho} s | {B}_s \rangle
+ \langle 0 | \bar{b} \overleftarrow{D}_{\rho} \overleftarrow{D}^{\rho} \gamma_{\mu} \gamma_5 s | {B}_s \rangle
+ \langle 0 | \bar{b} \gamma_{\mu} \gamma_5 D_{\rho} D^{\rho} s | {B}_s \rangle = -i f_{B_s} P_{\mu} M_{B_s}^2 e^{-iPx}.
\end{eqnarray}
For the latter one needs
\begin{eqnarray}
\label{DD}
&& D^2 = \not{\!\!D}\not{\!\!D} + \frac{g_s}{2} \sigma_{{\rho}{\nu}} F^{\rho\nu,a} T^a,
\end{eqnarray}
with the gluon field strength tensor $ F^{\rho\nu,a}$ in order to apply
$i \bar b \overleftarrow{\not{\!\!D}} =- m_b \bar b$ and
$i\!\!\not{\!\!\!D} s=
m_s s$, and $\sigma_{\rho\nu}=i(\gamma_{\rho}\gamma_{\nu}-\gamma_{\nu}\gamma_{\rho})/2$.
Using this relation we can write
\begin{eqnarray}
\label{A}
\langle 0 | \bar{b} \gamma_{\mu} \gamma_5 D_{\rho} D^{\rho} s | {B}_s \rangle
&=& -m_s^2 \langle 0 | \bar{b} \gamma_{\mu} \gamma_5 s | {B}_s \rangle
+ \frac{1}{2} \langle 0 | \bar{b} \gamma_{\mu} \sigma_{\rho\nu} g_s F^{\rho\nu,a} T^a \gamma_5 s | {B}_s \rangle,
\\
\label{B}
\langle 0 | \bar{b} \overleftarrow{D}_{\rho} \overleftarrow{D}^{\rho} \gamma_{\mu} \gamma_5 s | {B}_s \rangle
&=& -m_b^2 \langle 0 | \bar{b} \gamma_{\mu} \gamma_5 s | {B}_s \rangle
+ \frac{1}{2} \langle 0 | \bar{b} \sigma_{\rho\nu} \gamma_{\mu} g_s F^{\rho\nu,a} T^a \gamma_5 s | {B}_s \rangle.
\end{eqnarray}
Combining these results one arrives at
\begin{eqnarray}
 \langle 0 | \bar{b} \overleftarrow{D}_{\rho} \gamma_{\mu} \gamma_5 D^{\rho} s | {B}_s \rangle
&=& \frac{i}{2}f_{B_s} P_{\mu} e^{-iPx} \left(m_b^2+m_s^2-M_{B_s}^2\right)
- \frac{1}{2}\langle 0 | S_{\mu} | {B}_s \rangle.\label{rescurr}
\end{eqnarray}
where we have introduced the operator
\begin{eqnarray}
\label{oper1}
 S_{\mu} &\equiv& \frac{g_s}{2}\, \bar{b} F^{\rho\nu,a} T^a \{\gamma_{\mu}, \sigma_{\rho\nu}\} \gamma_5 s.
\end{eqnarray}
In \eq{rescurr} one can drop the term $m_s^2$ because it is of higher
order in our $1/m_b$ expansion. \eq{rescurr} holds in any
renormalization scheme. The terms with $\langle 0 | S_{\mu} | {B}_s
\rangle$  are formally of higher order in $1/m_b$, but in the $\ov{\rm
  MS}$ scheme  the power counting is not manifest. Just as in the case
of $R_2,\ldots \widetilde R_3$ in Sec.~\ref{rn} the matrix elements of
the current operator $S_{\mu}^{\ov{\rm MS}} $ contains unsuppressed
pieces, this time $\propto \alpha_s \langle 0 | \bar{b} \gamma_{\mu}
\gamma_5 s (x) | B_s \rangle$. Again, these terms result from hard loop
momenta, are IR finite and perturbatively calculable. Putting everything
together yields
\begin{eqnarray}
  m_b^2 \langle \bar B_s|{R}_2^{\ov{\rm MS}}|B_s\rangle^{\mathrm{large}\, N_c} &=& 
  - f_{B_s}^2 M_{B_s}^2 \left( M_{B_s}^2 - \bar m_b^2-\bar m_s^2 \right)
 - \frac{1}{2}   \langle
     \bar{B}_s | \bar{b} \gamma_{\mu} \gamma_5 s | 0 \rangle \langle 0 |
     S_{\mu}^{\ov{\rm MS}} | {B}_s \rangle \no\\
  && -  \frac{1}{2} \langle \bar B_s | S_{\mu} ^{\ov{\rm MS}}| 0 \rangle
     \langle 0  | \bar{b} \gamma_{\mu} \gamma_5 s | B_s \rangle . 
\label{factorR2n}
\end{eqnarray}
Dropping the $S_\mu$ terms, which are NLO in $\alpha_s$, reproduces the
result of the LO calculation in Ref.~\cite{Beneke:1996gn}. 

It remains to calculate the matrix element of $S_{\mu}$. The operator
$S_{\mu}$ defined in Eq. (\ref{oper1})  has the following Feynman rule 
\begin{eqnarray}
&& -ig_s k^{\rho} \{\gamma_{\mu},\sigma_{\rho\nu}\} T^a_{ij} \gamma_5,
\end{eqnarray}
where
$k$ is the incoming momentum of the gluon. A second coupling involves
two gluons and does not contribute at order $\alpha_s$.  Using this
Feynman rule we find the one-loop matrix element $S_{\mu}$ as%
\bea%
\label{Smu}
\langle 0 | S_{\mu}^{\ov{\rm MS}} | {B}_s \rangle^{(1)} &=& m_b^2\,
\frac{\alpha_s C_F}{4\pi}\,  4 \log \frac{m_b}{\mu} \, i f_{B_s}P_{\mu}e^{-iPx} +
\mathcal{O}(\alpha_s^2,m_b). %
\eea%
This result solely stems from hard loop momenta, is insensitive to the
external states and could therefore  calculated in perturbative QCD with external quark
states. At the calculated order the renormalization scheme for $m_b$ in
\eq{Smu} is not defined, this will matter only in an NNLO calculation. 

The calculation of Eq. (\ref{Smu}) has involved two steps: first the one-loop diagrams for $\langle S_{\mu}^{\overline{{\text{MS}}}} \rangle^{(1)}$ are determined in perturbation theory for free quarks as external states and expressed in terms of (tree-level) matrix elements of current operators. The pieces proportional to the $s$ quark momentum involve $\langle \bar{b}  \gamma_5 D_{\rho} s \rangle$. The second step to derive Eq. (\ref{Smu}) requires the determination of the hadronic matrix element $\langle 0 | \bar{b}  \gamma_5 D_{\rho} s |  {B}_s \rangle$, which we elaborate on here: starting with $D_{\rho} = \{ \gamma_{\rho},\not{\!\!D} \}/2$ we find
\begin{eqnarray}
\label{eqDrho}
  \langle 0 | \bar{b}  \gamma_5 D_{\rho} s |  {B}_s \rangle
  &=& \frac12  \langle 0 | \bar{b}  \gamma_5 \not{\!\!D} \gamma_{\rho} s |  {B}_s \rangle
          - \frac12  i m_s \langle 0 | \bar{b}  \gamma_5 \gamma_{\rho} s
      |  {B}_s \rangle \no\\
  &=&  \frac12  \partial_\mu
      \langle 0 | \bar{b}  \gamma_5 \gamma_{\mu} \gamma_{\rho} s |  {B}_s \rangle
      - \frac12  
      \langle 0 | \bar{b}  \gamma_5 \overleftarrow{\not{\!\!D}}
      \gamma_{\rho} s |  {B}_s \rangle  +
      \frac12  i m_s \langle 0 | \bar{b}   \gamma_{\rho} \gamma_5 s
      |  {B}_s \rangle
     \no\\
  &=&  \frac12  \partial_\mu
      \langle 0 | \bar{b}  \gamma_5 \gamma_{\mu} \gamma_{\rho} s |  {B}_s \rangle
      - \frac12 i (m_b-m_s)   
      \langle 0 | \bar{b}   \gamma_{\rho} \gamma_5 s
      |  {B}_s \rangle.
\end{eqnarray}
Using $\gamma_{\mu} \gamma_{\rho} = g_{\mu\rho} - i \sigma_{\mu\rho}$
and $ \langle 0 | \bar{b}  \gamma_5 \sigma_{\mu\rho} s |  {B}_s
\rangle=0$ one arrives at
\begin{eqnarray}
  \label{eqbds}
  \langle 0 | \bar{b}  \gamma_5 D_{\rho} s |  {B}_s \rangle
  &=&  -\frac12 f_{B_s} P_\rho \frac{M_{B_s}^2}{m_b+m_s} e^{-iPx}
      + \frac12 (m_b -m_s)  P_\rho f_{B_s} e^{-iPx} \no \\
  &=& - \frac12  f_{B_s} P_\rho \frac{M_{B_s}^2-m_b^2 + m_s^2}{m_b+m_s}e^{-iPx}.
\end{eqnarray}
{To specify to the $\ov{\rm MS}$ scheme} we finally need
\begin{eqnarray}
  \bar m_b(\mu)
  &=& m_b^{{\rm pole}} \lt[ 1 + \frac{\alpha_s(\mu)}{4\pi} C_F
       \lt( - 4  + 6 \log \frac{m_b}{\mu}\rt) \rt] \label{eq:polemass}
\end{eqnarray}
which implies
\begin{eqnarray}
  \frac{M_{B_s}^2}{\bar m_b^2} - 1
  &=&  \lt[ \frac{M_{B_s}^2}{m_b^{{\rm pole}\,2}}
                                     - 1 \rt] \,+\,  \frac{M_{B_s}^2}{m_b^2}
              \frac{\alpha_s(\mu)}{4\pi} 2 C_F   \lt( 4  - 6 \log
      \frac{m_b}{\mu}\rt)  +\mathcal{O}(\alpha_s^2) \no\\
  &=&  \frac{\alpha_s(\mu)}{4\pi} 2 C_F   \lt( 4  - 6 \log
      \frac{m_b}{\mu}\rt) +\mathcal{O}(\alpha_s^2,1/m_b). %
      \label{eq:rpol}
\end{eqnarray}  
{Using this in \eq{eqbds} we obtain the final result for the current
  matrix element
\begin{eqnarray}
    \label{eqbdsres}
  \langle 0 | \lt[ \bar{b}  \gamma_5 D_{\rho} s\rt]^{{\ov{\rm MS}}} |  {B}_s \rangle
  &=& - f_{B_s} m_b P_\rho e^{-iPx} \,
       \frac{\alpha_s(\mu)}{4\pi} C_F   \lt( 4  - 6 \log
      \frac{m_b}{\mu}\rt) +\mathcal{O}(\alpha_s^2,1/m_b).
\end{eqnarray}
Here the factor in brackets is specific to the $\ov{\rm MS}$ scheme,
while the the scheme of $m_b$ is still arbitrary, because a scheme
transformation will affect the result at order $\alpha_s^2$. $m_s$ has
been neglected, since terms of order $m_s/m_b$ are power-suppressed.}  
  
Recalling $C_F\simeq N_c/2$ and plugging Eq. (\ref{Smu}) into
Eq. (\ref{factorR2n}) we find
\begin{eqnarray}
\label{resR2-2}
  \langle \bar B_s|{R}_2^{\ov{\rm MS}}|B_s\rangle^{\mathrm{large}\,N_c}
  &=&
      f_{B_s}^2M_{B_s}^2 \,
      \frac{\alpha_sN_c}{4\pi}\left(4\log \frac{m_b}{\mu} - 4\right)
      +\mathcal{O}(\alpha_s^2,1/m_b).
\end{eqnarray}
Comparing this result with Eq. (\ref{resR2CF}) we see that the two
results are in agreement with each other.  Thus the properly
renormalized matrix elements satisfy 
$\langle  \bar B_s | {R}_2^{\rm ren}| B_s\rangle^{\mathrm{large}\,N_c}=\mathcal{O}
(1/m_b)$ at NLO, because the $\alpha_s$ term in \eq{resR2-2} is canceled by
$\delta R_2$.

\subsection{Operator $R_3$}
From \eq{eq:r3ms} one finds with \eq{eq:facme}:%
\bea%
\label{R3Nc} 
\langle  \bar B_s|R_3^{\ov{\rm MS}} | B_s \rangle^{\mathrm{large}\, N_c}
&=&  3 f_{B_s}^2M_{B_s}^2 \frac{\alpha_s\, N_c}{4\pi}
+\mathcal{O}(\alpha_s^2,1/m_b) .     %
\eea%
The large-$N_c$ matrix element of $R_3$ reads
\begin{eqnarray}
\label{factR3}
  m_b^2 \langle{R}_3\rangle^{\mathrm{large}\,N_c}
  &=& \langle
      \bar{B}_s |
      \bar{b} \gamma_5 s
      | 0 \rangle 
      \langle 0 | \bar{b} \overleftarrow{D}_{\rho} \gamma_5 D^{\rho} s | {B}_s  \rangle
       +
     \langle \bar B_s  | \bar{b} \overleftarrow{D}_{\rho} \gamma_5
       D^{\rho} s | 0 \rangle \langle 0 | \bar{b} \gamma_5 s | B_s \rangle.
\end{eqnarray}
Repeating the steps shown in the previous section associated operator
$R_2$, we get the following factorized formula (similar to
\eq{factorR2n}) for the color enhanced part of $R_3$
\begin{eqnarray}
  m_b^2 \langle \bar B_s|{R}_3^{\ov{\rm MS}}|B_s\rangle^{\mathrm{large}\, N_c}
  &=& \phantom{+}
      f_{B_s}^2 \frac{M_{B_s}^2}{(\bar m_b+\bar m_s)^2}
      \left( M_{B_s}^2 - \bar m_b^2-\bar m_s^2 \right)
      \no
      \\
  &&  - \frac{1}{2}   \langle
      \bar{B}_s | \bar{b} \gamma_5 s | 0 \rangle \langle 0 | G_S^{\ov{\rm MS}} | {B}_s \rangle
      -  \frac{1}{2} \langle \bar B_s | G_S^{\ov{\rm MS}}| 0 \rangle
     \langle 0  | \bar{b}  \gamma_5 s | B_s \rangle ,
\label{factorR3n}
\end{eqnarray}
where we have introduced the operator
\begin{eqnarray}
&& G_S = g_s\, \bar{b}_i F^{\rho\nu,a} T^a_{ij} \sigma_{\rho\nu} \gamma_5 s_j,
\end{eqnarray}
with the following Feynman rule
\begin{eqnarray}
&& - 2ig_s k^{\rho} \sigma_{\rho\nu} T^a_{ij} \gamma_5 ,
\end{eqnarray}
where $k$ is the incoming gluon momentum. At one-loop level
the matrix element of $G_S$  is found as
\begin{eqnarray}
  && \langle 0 | G_S^{\ov{\rm MS}} | {B}_s \rangle^{(1)} =
     -m_b^2 \frac{\alpha_sC_F}{4\pi}\cdot \left[- 2
     +
     12 \log \frac{m_b}{\mu} \right] if_{B_s}
     \frac{M_{B_s}^2}{m_b+m_s} e^{-iPx}+
     \mathcal{O}(\alpha_s^2,m_b). \label{GS} %
\end{eqnarray}
Also here the renormalization scheme for $m_{b,s}$ will only be fixed if one extends the calculation to NNLO.
Inserting \eqsand{eq:rpol}{GS} into \eq{factorR3n} gives 
\begin{eqnarray}
 \langle \bar B_s|{R}_3|B_s\rangle^{\mathrm{large}\,N_c}
  &=&
      -f_{B_s}^2M_{B_s}^2
      \frac{\alpha_sN_c}{4\pi}\left(6\log
      \frac{m_b}{\mu}-4\right)
      \no\\ 
  &&  + f_B^2 M_{B_s}^2 \frac{\alpha_sN_c }{4\pi}
     \left(6 \log \frac{m_b}{\mu} -1\right) +\mathcal{O}(\alpha_s^2,1/m_b) 
  \no \\
  &=& \phantom{+} 3 f_{B_s}^2M_{B_s}^2 \frac{\alpha_sN_c}{4\pi} +\mathcal{O}(\alpha_s^2,1/m_b) .     %.
\end{eqnarray}
Comparing this result with (\ref{R3Nc}) we verify that they are in
agreement with each other and thereby confirm $\langle  \bar B_s |
{R}_3^{\rm ren}| B_s\rangle^{\mathrm{large}\,N_c}=\mathcal{O} (1/m_b)$
at NLO.

\subsection{Operator $\widetilde R_2$}
From \eqsand{eq:r2tms}{eq:facme}  we get%
\bea%
\label{R2CNc}
\langle  \bar B_s| \widetilde R_2^{\ov{\rm MS}} | B_s \rangle^{\mathrm{large}\, N_c}
& = & 
f_{B_s}^2M_{B_s}^2 \frac{\alpha_s\,N_c}{4\pi} \cdot \left[ - 2 \log
  \frac{m_b}{\mu}+\frac{10}{3}
  -\frac{\widetilde{a}_2}{12}+\frac{\widetilde{b}_2}{6} \right]
{+\mathcal{O} (\alpha_s^2,1/m_b)} .%
\eea%
The large-$N_c$ limit of $\langle  \bar B_s|\widetilde R_2^{\ov{\rm MS}}
| B_s \rangle^{\mathrm{large}\, N_c}$ involves the Fierz transform of
$\widetilde R_2$:
\begin{eqnarray}
\label{R2CF}
 \widetilde{R}_2^{F} &\equiv& \frac{1}{m_b^2} \left(\bar{b}\overleftarrow{D}_{\rho}\right)_{\alpha}
\gamma_{\mu} L s_{\alpha} \otimes \bar{b}_{\beta} \gamma^{\mu} L \left(D^{\rho} s\right)_{\beta}.
\end{eqnarray}
\begin{eqnarray}
\label{factR2t}
  m_b^2 \langle \bar B_s | \widetilde R_2| B_s \rangle^{\mathrm{large}\,N_c}
  &=&  \langle \bar{B}_s | \bar{b} \overleftarrow{D}_{\rho} \gamma_{\mu} \gamma_5 s | 0 \rangle
      \langle 0 | \bar{b}  \gamma_{\mu} \gamma_5
      D^{\rho} s | B_s \rangle +
      \langle \bar B_s  | \bar{b} \gamma_{\mu} \gamma_5
      D^{\rho} s | 0 \rangle \langle 0 |
      \bar{b} \overleftarrow{D}_{\rho}\gamma_{\mu} \gamma_5  s | B_s \rangle. \qquad
\end{eqnarray}
Owing to
\begin{eqnarray}
\nonumber \langle \bar{B}_s | \bar{b}\overleftarrow{D}_{\rho} \gamma_{\mu} \gamma_5 s | 0 \rangle &=& \partial_{\rho} \langle \bar{B}_s | \bar{b} \gamma_{\mu} \gamma_5 s | 0 \rangle - \langle \bar{B}_s | \bar{b} \gamma_{\mu} \gamma_5 D_{\rho} s | 0 \rangle
\\
&=& f_{B_s} P_{\rho} P_{\mu} e^{iPx}  - \langle \bar{B}_s | \bar{b} \gamma_{\mu} \gamma_5 D_{\rho} s | 0 \rangle,
\end{eqnarray}
we only need to calculate
\begin{eqnarray}
\label{projection}
&& \langle 0 | \bar{b} \gamma_{\mu} \gamma_5 D_{\rho} s | {B}_s \rangle
   = \left[aP_{\mu}P_{\rho}+bg_{\mu\rho}M_{B_s}^2\right]f_{B_s}
   e^{-iPx}. 
\end{eqnarray}
{The desired matrix element in \eq{factR2t} reads 
\begin{eqnarray}
\label{projection-extra}
  m_b^2 \langle \bar B_s | \widetilde R_2| B_s \rangle^{\mathrm{large}\,N_c}
  &=& 2 f_{B_s}^2 M_{B_s}^4 \, \lt[ (1-a-4 b) (a+b) +3 a b\rt]
\end{eqnarray}
when expressed in terms of the real parameters $a$ and $b$. To the order
considered in this paper we can neglect $-a-4b$ in the first
  bracket, because $a$ and $b$ are ${\cal O}(\alpha_s, 1/m_b)$.

To determine the coefficients $a$, $b$ we first contract with
$g^{\mu\rho}$ to find:
\begin{eqnarray}
\label{eq1}
&& a+4b = \frac{m_s}{m_b+m_s},
\end{eqnarray}
which shows that this combination receives no radiative
  corrections and is further suppressed by a factor of $m_s/\lqcd$
  w.r.t.\ the nominal size of dimension-7 terms.
{The matrix element in \eq{projection} depends on the renormalization
  scheme employed for the operator and thus $a$ and $b$ are
  scheme-dependent.}  Now we apply $\partial_{\rho}$ to
Eq. (\ref{projection}), it gives
\begin{eqnarray}
\label{projection1}
&& \langle 0 | \bar{b} \overleftarrow{D}_{\rho} \gamma_{\mu} \gamma_5 D^{\rho} s | {B}_s \rangle
+ \langle 0 | \bar{b} \gamma_{\mu} \gamma_5 D_{\rho} D^{\rho} s | {B}_s \rangle = -P_{\rho} \left[aP_{\mu}P_{\rho}+bg_{\mu\rho}M_{B_s}^2\right] i f_{B_s} e^{-iPx}.
\end{eqnarray}
To solve this equation we use Eq. (\ref{factorR2}) and find
\begin{eqnarray}
&& \langle 0 | \bar{b} \overleftarrow{D}_{\rho} \gamma_{\mu} \gamma_5 D^{\rho} s | {B}_s \rangle
= \frac{1}{2}\left(-i f_{B_s} P_{\mu} M_{B_s}^2 e^{-iPx} - \langle 0 | \bar{b} \overleftarrow{D}_{\rho} \overleftarrow{D}^{\rho} \gamma_{\mu} \gamma_5 s | {B}_s \rangle
- \langle 0 | \bar{b} \gamma_{\mu} \gamma_5 D_{\rho} D^{\rho} s | {B}_s \rangle\right). \quad\quad
\end{eqnarray}
Using this result we make substitution in Eq. (\ref{projection1}) and get
\begin{eqnarray}
\label{projection2}
&& \langle 0 | \bar{b} \overleftarrow{D}_{\rho} \overleftarrow{D}^{\rho} \gamma_{\mu} \gamma_5 s | {B}_s \rangle
- \langle 0 | \bar{b} \gamma_{\mu} \gamma_5 D_{\rho} D^{\rho} s | {B}_s \rangle = \left[2a+2b-1\right] i f_{B_s} P_{\mu} M_{B_s}^2e^{-iPx}.
\end{eqnarray}
Also from Eqs. (\ref{A}) and (\ref{B}) we have
\begin{eqnarray}
&& \nonumber \langle 0 | \bar{b} \overleftarrow{D}_{\rho} \overleftarrow{D}^{\rho} \gamma_{\mu} \gamma_5 s | {B}_s \rangle
- \langle 0 | \bar{b} \gamma_{\mu} \gamma_5 D_{\rho} D^{\rho} s | {B}_s \rangle
\\
&& \nonumber \hspace{1.1cm} = -(m_b^2-m_s^2) \langle 0 | \bar{b} \gamma_{\mu} \gamma_5 s | {B}_s \rangle
+ \frac{1}{2} \langle 0 | \bar{b} [\sigma_{\rho\nu}, \gamma_{\mu}] g_s F^{\rho\nu,a} T^a \gamma_5 s | {B}_s \rangle
\\
&& 
\hspace{1.1cm} = -(m_b^2-m_s^2) if_{B_s}P_{\mu}e^{-iPx} + \langle 0 | \widetilde{S}_{\mu} | {B}_s \rangle,
\end{eqnarray}
where we have introduced
\begin{eqnarray}
&& \widetilde{S}_{\mu} = \frac{g_s}{2} \bar{b} F^{\rho\nu,a} T^a [\sigma_{\rho\nu}, \gamma_{\mu}] \gamma_5 s.
\end{eqnarray}
Inserting this result into Eq. (\ref{projection2}) we get
\begin{eqnarray}
\label{projection3}
&& -(m_b^2-m_s^2) if_{B_s}P_{\mu}e^{-iPx} + \langle 0 | \widetilde{S}_{\mu} | {B}_s \rangle = \left[2a+2b-1\right] i f_{B_s} P_{\mu} M_{B_s}^2e^{-iPx}.
\end{eqnarray}
From $\widetilde{S}_{\mu}$ we have the following Feynman rule for its one-gluon coupling
\begin{eqnarray}
&& -ig_s k^{\rho} [\sigma_{\rho\nu}, \gamma_{\mu}] T^a_{ij} \gamma_5,
\end{eqnarray}
where $k$ is an incoming momentum of a gluon. At one loop, using for outgoing anti-$b$ quark $\bar{b}\not{\!p}_b=-m_b\bar{b}$, we have
\begin{eqnarray}
\label{M1}
\langle 0 | \tilde{S}_{\mu}^{\ov{\rm{MS}}} | {B}_s \rangle^{(1)} = \frac{\alpha_sC_F}{4\pi} m_b^2 \left( - 2 + 8 \log \frac{m_b}{\mu}\right) if_{B_s} P_{\mu} e^{-iPx} + \mathcal{O}\left(\alpha_s^2,1/m_b\right).
\end{eqnarray}
Inserting this result into Eq. (\ref{projection3}) we obtain
\begin{eqnarray}
  && \nonumber -({\bar m}_b^2-{\bar m}_s^2) if_{B_s}P_{\mu}e^{-iPx} + \frac{\alpha_sC_F}{4\pi}m_b^2 \left( - 2 + 8 \log \frac{m_b}{\mu}\right) if_{B_s} P_{\mu} e^{-iPx} + \mathcal{O}\left(\alpha_s^2,1/m_b\right)
  \\
  && \hspace{3cm} = \left[2a^{\ov{\rm MS}} + 2b^{\ov{\rm MS}} - 1\right] i f_{B_s} P_{\mu} M_{B_s}^2e^{-iPx},
\end{eqnarray}
{where we have added superscripts to $a$ and $b$
  {as well as  bars to the masses} to clarify that the
  result is specific to the $\ov{\rm MS}$ scheme.}
{Next} we find
\begin{eqnarray}
\label{eq2}
  a^{\ov{\rm MS}} + b^{\ov{\rm MS}}
  &=&
      \frac{1}{2}\left(1 - \frac{{\bar m}_b^2-{\bar m}_s^2}{M_{B_s}^2} +
      \frac{\alpha_sC_F}{4\pi} \frac{m_b^2}{M_{B_s}^2} \left( - 2 + 8 \log \frac{m_b}{\mu}\right)\right) + \mathcal{O}\left(\alpha_s^2,1/m_b\right).
\end{eqnarray}
{Dropping $\bar m_s^2$ and using \eq{eq:polemass} this leads to
\begin{eqnarray}
\label{eq2b}
  a^{\ov{\rm MS}} + b^{\ov{\rm MS}}
  &=&
      \frac{1}{2}\left(1 - \frac{m_b^{\rm pole\,2}}{M_{B_s}^2} +
      \frac{\alpha_sC_F}{4\pi} \frac{m_b^2}{M_{B_s}^2}
      \left( 6 - 4 \log \frac{m_b}{\mu}\right)\right)
      + \mathcal{O}\left(\alpha_s^2,1/m_b\right).
\end{eqnarray}
Here the first two terms in the brackets combine to a term of order
$\lqcd/m_b$ and can be dropped, while the $\alpha_s$ piece is the
desired power-unsuppressed radiative correction. }
{By inserting \eq{eq2b} into \eq{projection-extra}}
we obtain
\begin{eqnarray}
\nonumber \langle\bar{B}_s|\widetilde{R}^{\ov{\rm MS}}_2|B_s\rangle^{\mathrm{large}~N_c} &=&
2 f_{B_s}^2 M_{B_s}^2\left(a^{\ov{\rm MS}} + b^{\ov{\rm MS}}\right) + \mathcal{O}\left(1/m_b\right)
\\
&=& f_{B_s}^2
    M_{B_s}^2\frac{\alpha_s}{4\pi}N_c\left(3-2\ln\frac{m_b}{\mu} \right)
    + \mathcal{O}\left(\alpha_s^2,1/m_b\right).
    \label{R2CNc-check}
\end{eqnarray}
Comparing this result with Eq.~(\ref{R2CNc}) we verify that they are in
agreement with each other and thereby confirm
$\langle \hspace{0.0cm} \bar B_s| \widetilde{R}_2^{\rm ren} | B_s
\rangle^{\mathrm{large}\, N_c}=\mathcal{O}(1/m_b)$ at NLO for
{the choices} $\widetilde{a}_2=12$ and $\widetilde{b}_2=4$
{found in Sec.~\ref{sec:eva}
  by imposing the requirement that Fierz symmetry is respected
  at the one-loop level}. {We remark that Fierz symmetry is a stronger
  constraint than consistency of \eqsand{R2CNc}{R2CNc-check}, which
  holds for any values of   $\widetilde{a}_2$ and  $\widetilde{b}_2$
  satisfying $\widetilde a_2- 2 \widetilde b_2=4$.}

\begin{table}
  \caption{Finite renormalization constants (see
    \eqsand{eq:rb}{eq:rtb}). The UV divergent constants $Z_{jQ}^{(11)}$,
    $Z_{j\widetilde Q_S}^{(11)}$, $\widetilde Z_{jQ}^{(11)}$, and
    $\widetilde Z_{j\widetilde Q_S}^{(11)}$ are 1/2 of the coefficients
    of $\log\mu$. {The large-$N_c$ analysis of Sec.~\ref{fc} has
      shown that a consistent result requires a Fierz-symmetric
      definition of the colour-flipped operators, namely
      $\widetilde{a}_2=12$, $\widetilde{b}_2=4$, and $\widetilde{b}_3=10$.}
    \label{tab}}
    \begin{displaymath}
\begin{array}{c@{~}|@{\hspace{3mm}}c@{\hspace{3mm}}|@{\hspace{3mm}}c@{\hspace{3mm}}|@{\hspace{3mm}}c}
  j &  0 &2 &3 \\\hline
    &&&\\[-3mm]
  Z_{jQ}^{(10)}
    &
      \ds       (N_c+1)  \log \frac{\mu}{m_b}
         & \ds \left(-\frac{4}{3}+\frac{5}{3 N_c}+\frac{10 C_F}{3}\right)
           \log \frac{\mu}{m_b}  -\frac{16}{9}
            &\ds
              - \left(\frac{1}{2}+\frac{1}{2 N_c}\right)
              \log \frac{\mu}{m_b}
  \\[4mm]
    &\ds +\frac{N_c^2+2 N_c -2}{N_c}
         &\ds -\frac{b_2-a_2}{24}+\left(\frac{17}{9}-\frac{a_2}{24}\right)\frac{1}{ N_c}+\frac{37 C_F}{9}
            &\ds - \left( \frac{1}{3}-\frac{b_3}{48} + \frac{3}{4N_c} +\frac{3C_F}{2}
              \right) 
  \\[3mm]\hline
    &&&\\[-3mm] 
  Z_{j\widetilde Q_S}^{(10)}
    &\ds  4 (N_c+1)
      \log \frac{\mu}{m_b}

         &\ds     \left(\frac{2}{3}+\frac{2}{3       N_c}+\frac{4
           C_F}{3}\right) \log \frac{\mu}{m_b} 
            &\ds  -\left(2+\frac{2}{N_c}\right) \log \frac{\mu}{m_b} 
  \\[4mm]
    &\ds +2 (N_c+1)
          &\ds    + \frac{2}{9} -\frac{b_2}{12}+
            \left(\frac{2}{9} -\frac{b_2}{12}\right) \frac{1}{N_c}- \frac{2 C_F}{9} 
            &\ds  - \left(\frac{13}{6}-\frac{b_3}{24}\right)\left(1+\frac{1}{N_c}\right) - 3 C_F 
  \\[3mm]\hline
  &&&\\[-3mm]
  \widetilde Z_{jQ}^{(10)}
    &\ds
         &\ds  \left(\frac{5}{3}-\frac{4}{3 N_c}-\frac{8 C_F}{3}\right) \log \frac{\mu}{m_b}
            &\ds - C_F \log \frac{\mu}{m_b}   \\[4mm]
    & &\ds +\frac{37}{18} - \frac{19}{9 N_c}-\left(\frac{32}{9}+\frac{\widetilde{b}_2-\widetilde{a}_2}{12}\right) C_F
            &\ds  - \frac{3}{4}
              - \left(\frac{2}{3}-\frac{\widetilde{b}_3}{24}\right)C_F
 \\[3mm]\hline
  &&&\\[-3mm]
  \widetilde Z_{j\widetilde Q_S}^{(10)}
    &\ds
         &\ds \left(\frac{2}{3}+\frac{2}{3
        N_c}+\frac{4 C_F}{3}\right) \log \frac{\mu}{m_b}
            &\ds - 4 C_F \log \frac{\mu}{m_b} -\frac{3}{2}
\\[4mm]
  &&\ds - \frac{1}{9} - \frac{1}{9 N_c} + \left(\frac{4}{9}-\frac{\widetilde{b}_2}{6}\right)C_F
         & \ds  -\frac{3}{2N_c} - \left(\frac{13}{3}-\frac{\widetilde{b}_3}{12}\right)C_F \\[-3mm]
\end{array}
\end{displaymath}
~\\[-2mm]
\hrule
\end{table}

\subsection{Operator $\widetilde R_3$}

From Eqs. (\ref{R3tMS}) and (\ref{eq:facme}) in the large-$N_c$ limit we have%
\bea%
\label{R3CNc}
\langle  \hspace{0.0cm} \bar B_s| \widetilde{R}_3^{\ov{\rm MS}} | B_s \rangle^{\mathrm{large}\, N_c}
&=& f_{B_s}^2M_{B_s}^2 \frac{\alpha_s\,N_c}{4\pi} \cdot \left[- 3 \log
  \frac{m_b}{\mu} + \frac{17}{6}-\frac{\widetilde{b}_3}{12} \right] + \mathcal{O}\left(\alpha_s^2,1/m_b\right).%
\eea%
The large-$N_c$ matrix element of operator $\widetilde{R}_3$ involves the Fierz transform of $\widetilde{R}_3$:
\begin{eqnarray}
&& m_b^2 \widetilde{R}_3^{F} = -\frac{1}{2}\left(\bar{b}\overleftarrow{D}_{\rho}\right)_{\alpha}
L s_\alpha \otimes \bar{b}_{\beta} L \left(D^{\rho} s\right)_{\beta}
-\frac{1}{8}\left(\bar{b}\overleftarrow{D}_{\rho}\right)_{\alpha}
\sigma_{\mu\nu} L s_\alpha \otimes \bar{b}_{\beta} \sigma^{\mu\nu} L \left(D^{\rho} s\right)_{\beta}.
\end{eqnarray}
For the color enhanced part of factorized $\widetilde{R}_3$ we get
\begin{eqnarray}
&& m_b^2 \langle  \hspace{0.0cm} \bar B_s| \widetilde{R}_3 | B_s \rangle^{\mathrm{large}\, N_c} = 
\\
&& \nonumber \hspace{2cm} -\frac{1}{2}\langle \bar{B}_s | \bar{b}\overleftarrow{D}_{\rho} \gamma_5 s | 0 \rangle \langle 0 | \bar{b} \gamma_5 D^{\rho} s | {B}_s \rangle
-\frac{1}{2} \langle \bar{B}_s |\bar{b} \gamma_5 D^{\rho} s | 0 \rangle \langle 0 | \bar{b}\overleftarrow{D}_{\rho} \gamma_5 s | {B}_s \rangle 
\\
&& \nonumber \hspace{2cm} -\frac{1}{8}\langle \bar{B}_s | \bar{b}\overleftarrow{D}_{\rho}
\sigma_{\mu\nu} \gamma_5 s  0 \rangle \langle 0 | \bar{b} \sigma^{\mu\nu} \gamma_5 D^{\rho} s | {B}_s \rangle
-\frac{1}{8} \langle  \bar{B}_s | \bar{b} \sigma^{\mu\nu} \gamma_5 D^{\rho} s | 0 \rangle \langle 0 | \bar{b}\overleftarrow{D}_{\rho} \sigma_{\mu\nu} \gamma_5 s_\alpha | {B}_s \rangle,
\end{eqnarray}
which we further modify to the following form:
\begin{eqnarray}
&& m_b^2 \langle  \hspace{0.0cm} \bar B_s| \widetilde{R}_3 | B_s \rangle^{\mathrm{large}\, N_c} = 
-\frac{1}{2} \left(\partial_{\rho}  \langle \bar{B}_s | \bar{b} \gamma_5 s | 0 \rangle - \langle \bar{B}_s | \bar{b} \gamma_5 D_{\rho} s | 0 \rangle \right) \langle 0 | \bar{b} \gamma_5 D^{\rho} s | {B}_s \rangle +{c.c.}
\\
&& \nonumber \hspace{2.7cm} - \frac{1}{8} 
\left(\partial_{\rho} \langle \bar{B}_s | \bar{b} \sigma_{\mu\nu} \gamma_5 s_\alpha | 0 \rangle - \langle \bar{B}_s | \bar{b} \sigma_{\mu\nu} \gamma_5 D_{\rho} s_\alpha | 0\rangle \right) \langle 0 | \bar{b} \sigma^{\mu\nu} \gamma_5 D^{\rho} s | {B}_s \rangle + {c.c.}\,.
\end{eqnarray}
We have
\begin{eqnarray}
  &&  \langle \bar{B}_s| \bar{b} \sigma_{\mu\nu} \gamma_5
      D_{\rho} s | 0 \rangle =0,
\end{eqnarray}
thus we obtain
\begin{eqnarray}
  \nonumber m_b^2 \langle \bar{B}_s|\widetilde{R}_3|
  B_s \rangle^{\mathrm{large}\, N_c}
  &=& -\frac{1}{2}
      \lt( \partial_{\rho} \langle \bar{B}_s |
      \bar{b}\gamma_5 s | 0 \rangle \rt)
      \langle 0 | \bar{b} \gamma_5
      D^{\rho} s | {B}_s \rangle +{c.c.}
  \\
  && \nonumber -\frac{1}{2} 
     \langle \bar{B}_s | \bar{b} \gamma_5 D_{\rho} s | 0 \rangle
     \langle 0 | \bar{b} \gamma_5 D^{\rho} s | {B}_s \rangle +{c.c.}
  \\
  && - \frac{1}{8} \langle \bar{B}_s | \bar{b} \sigma^{\mu\nu} \gamma_5 D^{\rho} s | 0 \rangle
     \langle 0 | \bar{b} \sigma_{\mu\nu} \gamma_5 D_{\rho} s_\alpha | {B}_s \rangle +{c.c.}\,.
\end{eqnarray}
{In} this expression the terms {in the last two lines are
  suppressed by powers of $\alpha_s$ or $\Lambda_{QCD}/m_b$ w.r.t.\ the
  first term and can be neglected.}
So using {\eqsand{matelS}{eqbdsres}} we get
\begin{eqnarray}
  \nonumber \langle \bar{B}_s| \widetilde{R}_3^{{\ov{\rm MS}}}|
  B_s \rangle^{\mathrm{large}\, N_c}
  &=& -\frac{1}{2m_b^2} P_{\rho} f_{B_s}
      \frac{M^2_{B_s}}{{\bar m}_b+{\bar m}_s}e^{iPx}
      \langle 0 | \bar{b} \gamma_5 D^{\rho} s | {B}_s \rangle +{c.c.}
  \\
  &=& f_{B_s}^2M_{B_s}^2 \frac{\alpha_sN_c}{4\pi} \left(2-3\ln\frac{m_b}{\mu}\right) 
      + \mathcal{O}\left(\alpha_s^2,1/m_b\right),
\end{eqnarray}
which is in agreement with Eq. (\ref{R3CNc}) for $\widetilde{b}_3=10$,
{and Fierz symmetry is necessary and sufficient for the correct
  result for $ \langle \widetilde{R}_3 \rangle^{\mathrm{large}\, N_c}$.}

\section{Discussion and conclusions\label{cn}} {Current data on the
  width difference $\dg_s$ in the $B_s$--$\bar B_s$ system are more
  precise than the theory prediction
  \cite{Gerlach:2022hoj,Gerlach:2025tcx} and the uncertainty of the
  latter is dominated by {the} size of the power-suppressed contribution
  $\widetilde\Gamma_{21,1/m_b}^{cc}$ in \eq{eq::Gam^ab} to the decay
  matrix. This calls for a calculation of NLO QCD corrections to this
  quantity, which is novel territory for the renormalization problem of
  four-quark operators. It is known that, generally, in the
  $\ov{\rm MS}$ scheme the decoupling of heavy scales is not manifest
  and higher-dimensional operators can mix into lower-dimensional ones
  under renormalization. At present, it is not known whether the
  commonly adopted HQE
  \cite{hqe,hqe2,hqe3,hqe4,Beneke:1996gn,Beneke:1998sy,
    Beneke:2003az,Ciuchini:2003ww,Lenz:2006hd,Gerlach:2021xtb,
    Gerlach:2022wgb,Asatrian:2017qaz,Gerlach:2022hoj,Gerlach:2025tcx}
  for $\Gamma_{21}^q$ can be consistently defined and, if yes, how the
  finite counterterms to the dimension-7 operators restoring the power
  counting must be defined.}

{In this work we have done a first step  in this direction
  and} calculated those parts of the $\overline{\rm MS}$-renormalized matrix
elements of the dimension-7 operators $R_{2,3}$ and
$\widetilde{R}_{2,3}$ at order $\alpha_s$, which are associated with the
finite renormalization of these operators {by terms proportional to
  dimension-6 operators. We have found that the loop contributions
    to $\langle R_{2,3} \rangle $ and
    $\langle \widetilde{R}_{2,3}\rangle$ scaling like $m_b^0$ are indeed
    infrared-finite, so that they can be cancelled by counterterms
    proportional to the dimension-6 operators $Q$ and $\widetilde Q_S$
    to restore the correct power counting,
    $\langle R_{2,3}^{\rm ren}\rangle, \langle
    \widetilde{R}_{2,3}^{\rm ren}\rangle =\mathcal{O} (1/m_b)$. However,
    we have also
    found that these counterterms depend on the definitions of
    evanescent operators and have shown that in the case of the
    colour-flipped operators $\widetilde{R}_{2,3}$ the $\epsilon$ terms
    of these operators must be chosen to obey Fierz symmetry. This
    feature has been observed from the hadronic matrix elements
    calculated in the large-$N_c$ limit, which turned out to be very
    subtle, with power-unsuppressed contributions from hard loop
    corrections to operators involving the gluon field strength tensor.
    The large-$N_c$ criterion does not apply to the colour-straight
    operators $R_{2,3}$ and it is still an open question, whether
    arbitrary choices of the $\epsilon$ terms in the evanescent operators
    are permissible or whether other conditions will fix them, for
    example
    $\langle R_2\rangle=-\langle \widetilde R_2\rangle + {\cal O}
    (1/m_b^2) $, which automatically holds at tree-level
    \cite{Beneke:1996gn} but needs a finite counterterm at NLO.}
  {Since $Z_{1Q}=Z_{1\widetilde Q_S}=0$ and $Z_{0Q}$, $Z_{0\widetilde
      Q_S}$ could be extracted from Ref.~\cite{Beneke:1998sy}, the
    calculations presented in this paper complete the set  of 
  renormalisation constants related to the mixing of $R_j$, $\widetilde R_j$
  into dimension-6 operators.}
    
In conclusion, we have made a first step towards the calculation of the
$B$-$\bar B$ decay matrix at order $\alpha_s/m_b$, \textit{i.e.}\ to
extend the 29-year-old LO result of Ref.~\cite{Beneke:1996gn} to NLO, by
calculating the counterterms listed in \tab{tab}.
{As a by-product, we have calculated matrix elements of several
$\bar b$-$s$ current operators. Our results for the hard,
infrared-finite loop contributions are needed to restore the correct power
counting in any application involving these currents.}

\section*{Acknowledgements}
We thank Martin Lang for pointing out mistakes in Eqs.~(4.16),
  (4.33), and (4.34) in an earlier version of this paper. While not
  affecting the results of this paper, the correct 
  formulae are needed in future calculations of corrections at order $\alpha_s^2/m_b$
  and $1/m_b^2$.
UN is grateful to Veronika Chobanova for clarifications concerning
Ref.~\cite{HeavyFlavorAveragingGroupHFLAV:2024ctg}.
The work of
A.H. has been supported by the State Committee of Science of Armenia
Program, Grant No. 21AG-1C084.  UN's research was supported by the
Deutsche Forschungsgemeinschaft (DFG, German Research Foundation) under
grant 396021762 - TRR 257 for the Collaborative Research Center
\emph{Particle Physics Phenomenology after the Higgs Discovery (P3H)}.

\end{document}